# Density of Quantum States in Quasi-1D layers

D. Kakulia,[a)] A. Tavkhelidze,[b),*] V. Gogoberidze,[a)] and M. Mebonia[b, c)]

[a] *Tbilisi State University, Chavchavadze Ave. 3, 0179 Tbilisi, Georgia*
[b] *Ilia State University, Cholokashvili Ave. 3-5, 0162 Tbilisi, Georgia*
[c] *JCNS and Semiconductor Nanoelectronics, Peter Grünberg Institut PGI, Forschungszentrum Jülich GmbH, D-52425 Jülich, Germany*
*Corresponding author e-mail address: avtotav@gmail.com*

Recently, new quantum effects have been studied in thin nanograting layers. Nanograting on the surface imposes additional boundary conditions on the electron wave function and reduces the density of states (DOS). When the nanograting dimensions are close to the de Broglie wavelength, the DOS reduction is considerable and leads to changes in the layer properties. DOS calculations are challenging to perform and are related to the quantum billiard problem. Performing such calculations requires finding the solutions for the time-independent Schrödinger equation with Dirichlet boundary conditions. Here, we use a numerical method, namely the method of auxiliary sources, which offers significant computational cost reduction relative to other numerical methods. We found the first five eigenfunctions for the nanograting layer and compared them with the corresponding eigenfunctions for a plain layer by calculating the correlation coefficients. Furthermore, the numerical data were used to analyze the DOS reduction. The nanograting is shown to reduce the probability of occupation of a particular quantum state, reducing the integrated DOS by as much as 4.1 -fold. This reduction in the DOS leads to considerable changes in the electronic properties.

## 1. Introduction

Developments in nanotechnology have enabled the fabrication of sub-10-nm densely packed periodic structures [1, 2]. Moreover, nanograting (NG) has been demonstrated to notably change the electronic [3], thermoelectric [4], and electron emission properties [5] of materials when the grating pitch becomes comparable with the de Broglie wavelength of the electron. Such changes are due to the special boundary conditions imposed by NG on the electronic wave function. Supplementary boundary conditions forbid some quantum states and hence reduce the density of states (DOS), i.e., the number of states per interval of energy at each energy level that are available to be occupied.

To achieve layers with predesigned properties, a calculation of the DOS dependence on the NG dimensions is necessary. However, performing such a calculation is challenging. Such a calculation requires solving the time-independent Schrödinger equation for a NG geometry. Mathematically, there is no difference between the DOS reduction and the electromagnetic transverse-magnetic (TM) mode depression [6-8]. The Helmholtz equation with Dirichlet boundary conditions is used in both cases. Approximate analytical techniques for DOS calculation include Weyl's formula [9, 10] and ray theory [11]. Various grating configurations were numerically studied in [12]. A relatively new numerical approach used for quantum systems is the electromagnetic analog circuit method [13, 14]. Related quantum systems, such as the narrow-wide-narrow geometry quantum wire [15], periodic curved surfaces [16, 17], electron waveguides [18], and strain-driven nanostructures [19], have also been investigated.

Here, we use a numerical method, the method of auxiliary sources (MAS) to analyze quantum systems with NG geometry. In general, the MAS yields a solution that is the sum of the fundamental solutions (referred to as auxiliary sources, or AS for short) of the relevant partial differential equation (the Helmholtz wave equation in this particular case). Usually, the singularities of the fundamental solutions are distributed on some contour that surrounds the area of interest. The amplitudes of the AS are defined by the boundary conditions.

Figure 1 shows a typical cross-section of a NG layer. The grating on the surface has depth $a$ and pitch $2w$. For comparison, we choose as a reference layer a planar layer with thickness $H$, such that it has the same cross-sectional area. The NG layer is a quasi-1D system in the range $0 < a < 2H$. The NG layer is a 2D quantum well for $a = 0$ and a system of 1D quantum wires for $a = 2H$. The NG does not change the quantum properties in the Z direction. Here, we assume that the NG pitch is much larger than a lattice constant and that miniband formation can be neglected.

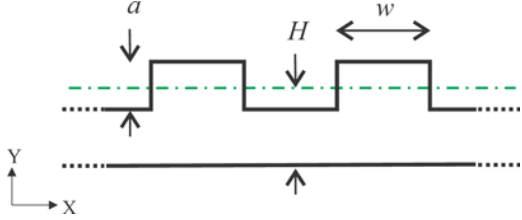

- Fig. 1. Cross-section of the nanograting layer.

Nanograting imposes additional boundary conditions on the wave function of the electron and reduces the number of quantum states that are available to be occupied. The DOS as a function of energy $\rho(E)$ diminishes with respect to that of the reference layer

$$\rho(E) = \rho_0(E) g(a, H, w, E). \qquad (1)$$

where $\rho_0(E) = \sum_n \delta(E - E_n)$ is the DOS for the reference layer, $g(E, H, w, a)$ is the probability mass function, $E_n$ is the energy for the n-th level of the reference quantum system (the planar layer in our case), and $\delta$ is the Dirac delta function. The function $g$ is introduced to describe the geometry-induced changes in the number of states that are available to be occupied by the electrons. For simplicity, we assume that $H = w = 1$ and $g = g(a, E)$. The physical meaning of $g(a, E)$ is the following: Consider an electron in a particular quantum state with wave vector $\vec{k}_n$ (and energy $E_n$) moving in space from point A to point B. Let this movement be possible only through two parallel paths; one path goes through the planar layer and the other through the NG layer (of depth $a$). If the probability of passing through the planar layer is $P$, then the probability of passing through the NG layer is $P \times g(a, E_n)$. This setup corresponds to an electron movement in a real device, very similar to that in an asymmetric Aharonov-Bohm ring [20] with half of the ring composed of a NG layer and the other half composed of a planar layer (Fig. 2). The introduction of $g = g(a, E)$ is convenient for calculating the electronic properties of the NG layers and comparing them with the planar layer. It should be noted, that $g(a, E)$ is layer material independent.

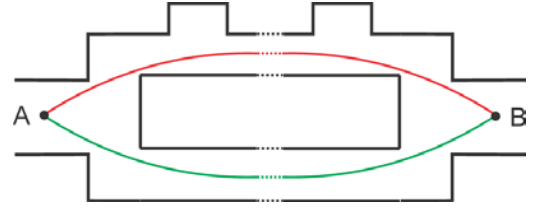

Fig. 2 Two possible paths of electron moving from point A to point B through asymmetric Aharonov-Bohm ring.

Main objective of this work is to calculate the function $g$ numerically and find a value for the grating depth $a$ for which the value of $g(E, a)$ has a minimum. Having this type of data is very important in the experimental study of NG layers.

The article is organized as follows. In section 2, we provide a detailed description of the MAS and the procedure to find eigenvalues and eigenfunctions. In section 3, we present eigenvalues and eigenfunctions of the time-independent Schrödinger equation for the first five quantum states in both the NG and the planar layers and build the spatial dependence of the probabilities $\varphi^2$, where $\varphi$ is the electron wave function. In section 4, we find $g(a)$ and use it to calculate the changes in the DOS and estimate the corresponding changes in the electronic properties. Conclusions are given in section 5.

An analysis is made within the limits of the approximations of a parabolic band and a wide quantum well. A parabolic band approximation can be used as we consider only band edges. A wide quantum well is a good approximation, as we regard relatively thick layers with better lateral transport properties.

## 2. Method of Auxiliary Sources

The MAS is a numerical method that is used mainly for electromagnetic problems [21-24]. The conceptual basis of the MAS lies between that for the method of moments and the method of images [23, 24]. The MAS has been justified mathematically by Kupradze [25], who proved the completeness and linear independence of the fundamental solutions of the Helmholtz equation when the poles are distributed over a closed "auxiliary" contour. Field sources (singularities of the fundamental solutions or the Green functions) are located on an auxiliary contour and are referred to as AS. The method can be applied to problems of acoustics and quantum mechanics, which are quantitatively similar to electromagnetic problems.

## 2.1 Technique to Obtain Eigenvalues and Eigenfunctions

We consider the eigenvalue problem for the two-dimensional (2D) Laplace operator over an area with periodic boundaries. For arbitrary-shaped boundaries, the problem must be solved numerically. Related methods include the finite element method and the boundary element method [26]. Here, we present a very simple and highly efficient numerical approach based on the MAS. The efficiency of this method was analyzed by Aleksidze [27].

To give a short description of how eigenvalues are found using the MAS, let us begin with a discussion of the 2D scattering problem. Consider a finite object placed in a primary scalar field. Let us find the scattered field outside of the object for Dirichlet boundary conditions (at the boundary, the sum of scattered and primary fields is zero). According to the MAS, the scattered field is represented by the sum of fields of the AS distributed on an auxiliary contour inside the object. The amplitudes of the AS can be found by satisfying the boundary conditions. Next, consider the same problem from the other side of the object's boundary. The object is hollow, the same primary field is defined, and the boundary conditions are the same; however, the auxiliary contour and the AS are located outside of the boundary. Following the path to the exact mathematical solution, the total field must be zero in a closed area, except those cases when $k^2$ in the Helmholtz equation, $\Delta \varphi + k^2 \varphi = 0$ (the time-independent Schrödinger equation in our case), is the eigenvalue of the Laplace operator, $-\Delta \varphi = k^2 \varphi$, where $k$ is the wave number and $\varphi$ is the total field (the electron wave function in our case), which can be written in the following form

$$\varphi(k,\vec{r}) = \sum_{i=1}^{N} \alpha_i \, G(k,\vec{r}_i',\vec{r}) + \varphi^{pri}(k,\vec{r}). \qquad (2)$$

Here, $G(k,\vec{r}_i',\vec{r})$ is the AS function (usually area Green function), $\alpha_i$ is the amplitude, $\vec{r}_i$ is the location vector of the $i$-th AS, $\vec{r}$ is the radius vector, $\varphi^{pri}(k,\vec{r})$ is a primary field, and $N$ is the number of AS.

At the beginning of the numerical evaluation, the incident field is only partly compensated for by the AS fields, and the total field is close to zero (not exactly zero). Compensation is improved by increasing $N$ and achieving convergence to the numerical solution. Such parameters as $N$ and the distance between the AS and boundaries strongly influence the result [24]. When the $k$ value approaches an eigenvalue, the total field inside the domain dramatically increases. By scanning over all $k$ values, we obtain the field curve with resonances near the eigenvalues. The resonance widths depend on the accuracy of fulfilling the boundary condition. When the boundary conditions are better satisfied, the resonances become narrower. Furthermore, $k$ converges to the eigenvalue $k_n$. Finally, for a fixed $k_n$, the eigenfunctions are found by scanning over the 2D domain.

## 2.2 Grouping of AS and Green's function

In the case of periodic boundaries, the AS can be grouped because they have periodically repeating amplitudes [21-24]. The Green function grouping gives the following sum

$$G(k,\vec{r}_i',\vec{r}) = \frac{j}{4} \sum_{i=-\infty}^{\infty} H_0^{(1)}\left(k\sqrt{(x-nd-x')^2 + (y-y')^2}\right). \qquad (3)$$

Here, $j$ is the imaginary unit, $H_0^{(1)}$ is the zero-order Hankel function of the first kind, $d$ is the period of the boundary, $i$ is an index of the grouped AS, and $x, y, x', y'$ are the components of the $\vec{r}, \vec{r}_i'$ vectors, respectively. Equation (3) must be inserted into Eq. (2) for all values of index $i$; the unknown amplitudes $\alpha_i$ can be found by ensuring that Eq. (2) satisfies the boundary conditions. Equation (3) converges very slowly, which is computationally time consuming. The Poisson summation has been applied to the sum in Eq. (3) to accelerate the summation process [24]. Lampe [28] catalogued the Poisson summations for the potential fields of the Helmholtz equation and developed a technique to accelerate this summation. Poisson summation converts Eq. (3) to the following expression:

$$G(k,\vec{r}_i',\vec{r}) = \frac{1}{2d} \sum_{i=-\infty}^{\infty} \frac{\exp\left[j(\vec{r}-\vec{r}_i')\vec{k}^m\right]}{k_y^m} \qquad (4)$$

where $\vec{k}^m = (k_x^m, k_y^m)$ is a complex vector in 2D, $k_x^m = \frac{2\pi m}{d}$, and $k_y^m = sign\,(y-y')\sqrt{k^2 - (k_x^m)^2}$.

## 2.3 Numerical calculations

The numerical calculations were performed using the MATLAB program. Some 200 of the evenly distributed AS were placed along the boundary of an open 2D area at a distance $\varepsilon$ from the boundary (Fig. 3).

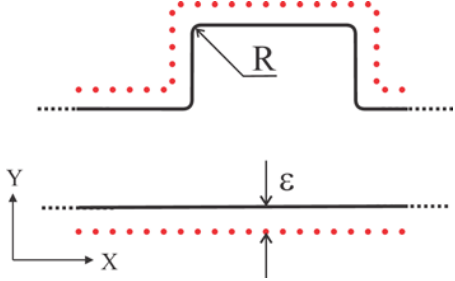

Fig. 3. One period of the NG area and the corresponding auxiliary sources.

The angles of the boundary were rounded, resulting in smooth contours. Such minute changes in the geometry influence the values of only very high-order eigenvalues and eigenfunctions. The radius of curvature $R$ was much less than the structural dimensions $R \ll a, w, H$. $R$ was also less than the distances between AS and the boundary $R < \varepsilon$. The eigenvalues and eigenfunctions were calculated over a 2D area covering one period of NG. The eigenvalues and eigenfunctions change when $a, H$, and $w$ are varied. For simplicity, we set $w = H = 1$ and vary the $a/H$ ratio by changing $a$. The area was divided into $N = 3761$ (61 × 61) parts, and the values for the eigenfunction were found for each part. The reference area was chosen as a plane area ($a = 0$). The correlation between the eigenfunctions of the NG area and the corresponding eigenfunctions of the reference area ($a = 0$) was considered. This process was performed because supplementary boundary conditions are the internal variables ($a$ in our case) of the eigenfunctions. As a result, the eigenfunction for a particular geometry ($a \neq 0$) is correlated with the corresponding eigenfunction of the reference geometry ($a = 0$).

The correlation coefficients between the squares of the eigenfunctions (the probabilities $\phi^2$ in our case) were calculated using the following discrete formula:

$$C_{0,a} = \sum_{p=1}^{N} \left(q_p^{(0)} - \overline{q^{(0)}}\right)\left(q_p^{(a)} - \overline{q^{(a)}}\right) / $$
$$/ \sqrt{\sum_{p=1}^{N} \left(q_p^{(0)} - \overline{q^{(0)}}\right)^2 \left(q_p^{(a)} - \overline{q^{(a)}}\right)^2} \quad (5)$$

where $q_p^{(0)} = (\varphi_p^{(0)})^2$, $q_p^{(a)} = (\varphi_p^{(a)})^2$, $\varphi_p^{(0)}$ and $\varphi_p^{(a)}$ are the values of the eigenfunctions in the reference and the NG geometries; $\overline{q^{(0)}} = \frac{1}{N}\sum_{p=1}^{N}(\varphi^{(0)})^2$ and $\overline{q^{(a)}} = \frac{1}{N}\sum_{p=1}^{N}(\varphi^{(a)})^2$ are the corresponding average values.

## 3. Eigenvalues and Eigenfunctions for the NG Geometry

The first five eigenvalues $k_n$ (n = 1, .., 5) were calculated for six different geometries $a = 0, 0.1, \ldots, 0.5$, resulting in 30 eigenvalues. Numerically calculated eigenvalues for the reference area ($a = 0$) were compared with well-known analytical solutions (πn) to verify the program code. The numerical results show that the eigenvalues for the NG areas deviate from those for the reference area. The dependence of deviation on NG geometry is shown in Fig4.

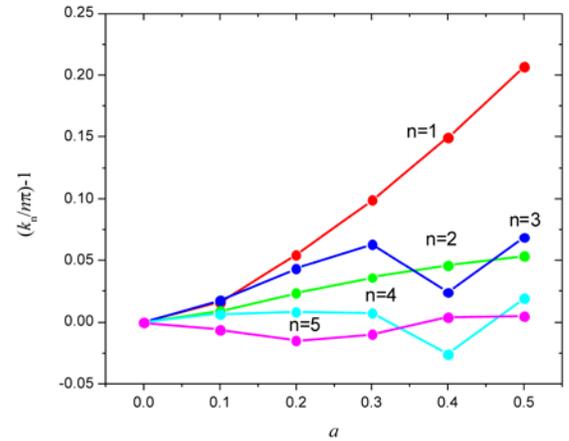

Fig. 4. Dependence of the eigenvalues on the nanograting depth for the first five eigenvalues.

As Fig. 4 shows, the eigenvalues increase with increasing $a$. Moreover, the eigenvalues oscillate for n>2. These oscillations can be explained by the presence of the well-known oscillatory part in the DOS [29, 30]. The shift of the eigenvalue towards higher values was significant only for n=1.

The eigenfunctions were calculated over the entire 2D area for all 30 eigenvalues. These eigenfunctions were evaluated by the following procedure. For a particular quantum state, the reference area eigenfunction $\varphi_p$ (the electron wave function in our case) was found at all $p$ points of the 2D area. Next, the correlation coefficients $C_{(0,a)}^{(n)}$ were calculated using Eq. (5). Figure 5 shows the correlation coefficient dependence on geometry for all five quantum states (n = 1, .., 5). A table containing the full set of $C_{(0,a)}^{(n)}$ values is given in the supplementary material [31].

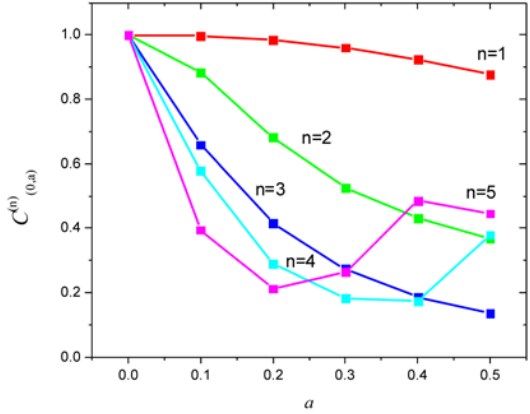

Fig. 5. Dependence of the correlation coefficients on geometry for five quantum states.

Figure 5 demonstrates that for low values of *a*, the correlation coefficient diminishes with increasing *a* (for all *n*). Furthermore, the correlation coefficient begins to increase again for *n*=4, 5. The minimum for a given $C_{(0,a)}^{(n)}(a)$ shifts to lower *a* values as *n* increases. This behavior can be explained by the following mechanism. Generally, the introduction of the supplementary boundary conditions (NG) reduces the correlation coefficient. This reduction is more pronounced for NG dimensions that are comparable with the de Broglie wavelengths ($2\pi/k_n$). Furthermore, Fig. 5 shows that the condition $a = \pi/k_n$ approximately holds for the minimum of the correlation coefficient, at least for *n*=4 and *n*=5. Condition $a = \pi/k_n$ corresponds to a plain de Broglie wave acquiring a phase difference $\pi$ (that is necessary for destructive interference) over distance 2*a*. For higher values of *a*, the NG dimensions surpass the de Broglie wavelength and the influence of the NG weakens, resulting in an increase in the correlation coefficient value.

The eigenfunctions were normalized in two stages. First, each eigenfunction was normalized by equating the entire probability of finding an electron inside the 2D area to unity, $\sum_{p=1}^{N}(\varphi_p)^2 = 1$. Next, the probability density obtained $(\varphi_p^{(a)})^2$ was normalized using the correlation coefficients $C_{(0,a)}$ as normalization constants according to $\sum_{p=1}^{N}(\varphi_p^{(a)})^2 = C_{(0,a)}\sum_{p=1}^{N}(\varphi_p^{(0)})^2 = C_{(0,a)}$. The obtained probabilities correspond to five electrons being in the first five quantum states inside the reference area. The normalization was performed to replicate the reduction in probability of finding the electron in a particular quantum state of area $a \neq 0$ with respect to the reference area. Figure 6 shows the resulting spatial distribution of the probability density for the fourth quantum state (*n*=4) for different values of $a$. The displayed X interval is shifted relative to the NG edge (as in Fig. 3) for clarity. Similar dependences for other values of *n* are presented in the supplementary material [31]. Figure 5 shows that the introduction of NG changes the shapes of the probability densities so that minima appear around the edge of the NG. The minima become more pronounced as *a* increases. These minima divide the peaks of the probability densities into parts that are small enough to fit inside the NG protrusion. For *a*=0.2 and 0.3, there are four peaks that fit inside this protrusion and only three peaks that fit inside the rest of the area. Some of the peaks correspond to fitting the electron inside the protrusion. The eigenfunctions of other quantum states behave qualitatively in the same manner upon introducing the grating. The peaks divide, and the minima appear close to the NG edge.

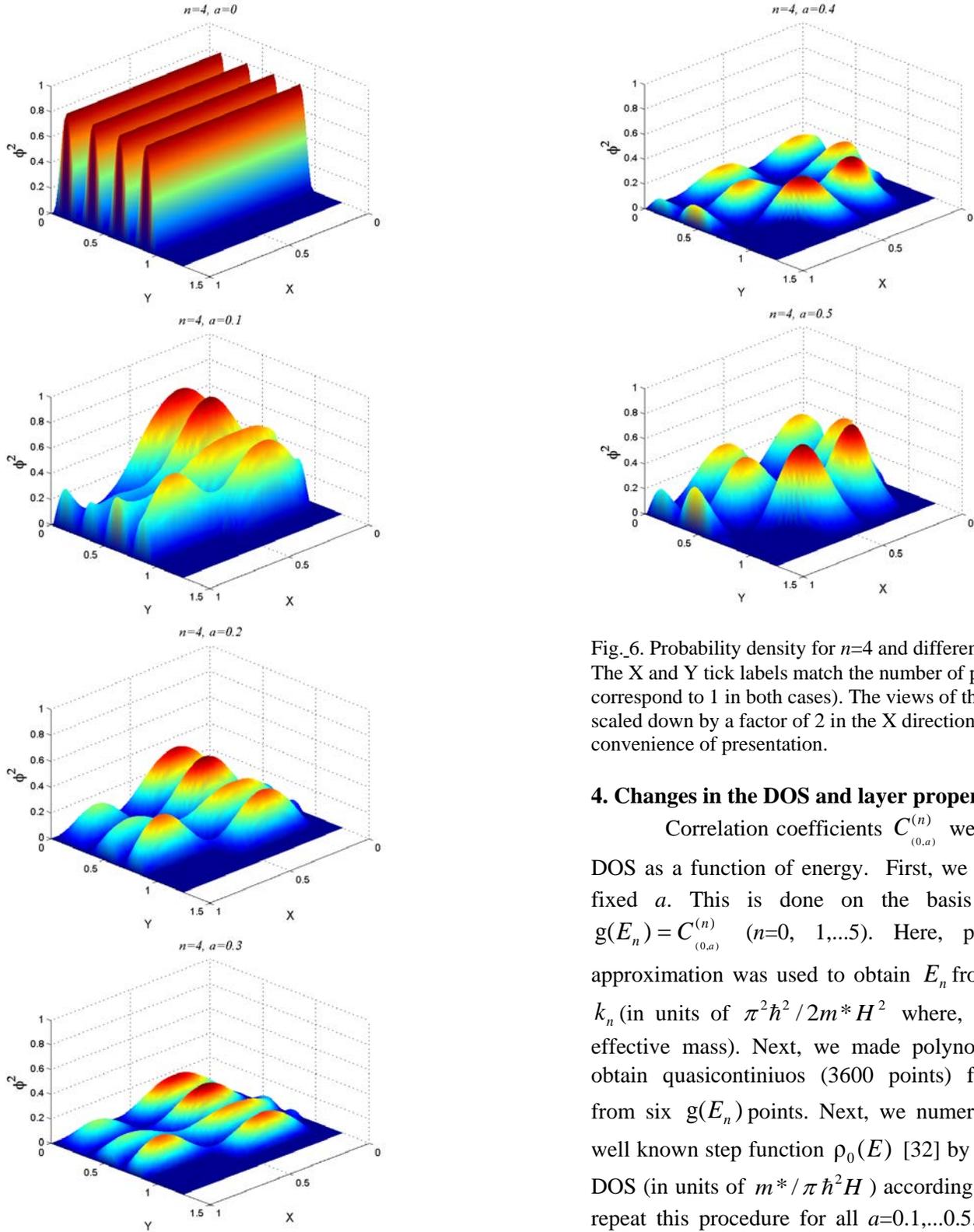

Fig. 6. Probability density for $n=4$ and different values of $a$. The X and Y tick labels match the number of pixels (61 pixels correspond to 1 in both cases). The views of the plots are scaled down by a factor of 2 in the X directions for convenience of presentation.

## 4. Changes in the DOS and layer properties

Correlation coefficients $C_{(0,a)}^{(n)}$ were used to find DOS as a function of energy. First, we build $g(E)$ for fixed $a$. This is done on the basis of 6 values $g(E_n) = C_{(0,a)}^{(n)}$ ($n=0, 1,...5$). Here, parabolic band approximation was used to obtain $E_n$ from eigenvalues $k_n$ (in units of $\pi^2\hbar^2/2m^*H^2$ where, $m^*$ is electron effective mass). Next, we made polynomial fitting to obtain quasicontiniuos (3600 points) function $g(E)$ from six $g(E_n)$ points. Next, we numerically multiply well known step function $\rho_0(E)$ [32] by $g(E)$ to obtain DOS (in units of $m^*/\pi\hbar^2 H$) according to Eq. (1). We repeat this procedure for all $a=0.1,...0.5$. The resulting spectrums are shown in Fig 7.

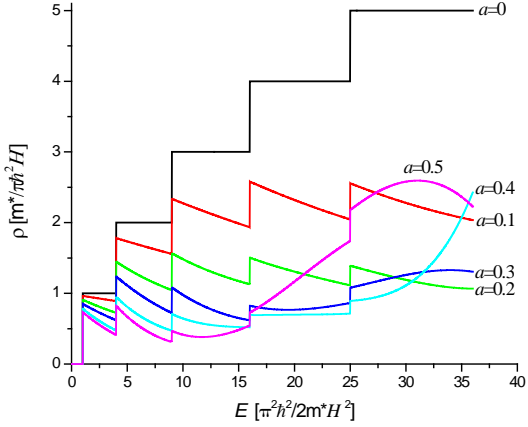

Fig. 7. DOS as a function of energy for different *a* values. Here, m* is electron effective mass.

As Fig. 7 shows, DOS reduce for all *a* values ower the considered energy region. This reduction is more significant for high *a* values. Last can be expleined by imposing of additional boundary conditions which are more pronounced for high *a* values. However, DOS increase back in high *E* region and for high *a* values (curves *a*=0.4, 0.5). This can be explained in terms of interference. NG strongly influences only eigenfunctions with low $k$ (low $E$) or with large de Broglie wavelengths $2\pi/k$ that are comparable to *a*. Eigenfunctions with higher $k$ values are slightly influenced by the NG, and the changes in DOS in high $E$ region become less significant. As Fig. 7 shows, spectrums are akin to 1D system [32] in the low $E$ region (especially for high *a*). This is natural as nanograting becomes more 1D like with increasing *a*.

Obtained data allows to find $g(a,E)$. Color contour of $g(a,E)$ is given in [31] and can be used to find its minimum. However, from the point of view of applications, it is more important to find minimum of integrated DOS. For this, we numerically integrate spectrums shown in Fig. 7 over the considered energy region. Next, we plot dependence $\overline{g}(a) = \int \rho_a(E)dE / \int \rho_0(E)dE$ as a function of *a* in Fig. 8. Here, the electron effective mass value m*=m was used (m is free electron mass).

As Fig. 8 shows $\overline{g}(a)$ has a minimum at *a*=0.4. Minimum value is $\overline{g}(0.4)$=0.24.

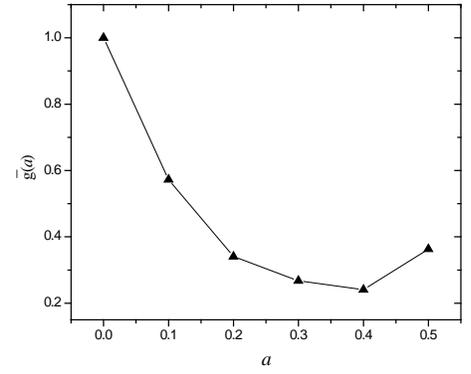

Fig. 8 $\overline{g}(a)$ as fuction of *a* for m*=m and *H*=1.

In general, using data from a limited number of quantum states gives only an approximate value of the DOS. However, for some common experimental designs, data for a few quantum states are sufficient. Consider a semiconductor quantum well layer grown on a semiconductor substrate. Only electrons from the first few energy levels have energies less than the band offset and are confined to the quantum well. In practice, the average distance between the energy levels of the semiconductor quantum well are on the order of a few meV and the band offset is a few tenths of a meV [32]. Consequently, data from the first 5 energy levels are sufficient for the DOS calculation in the energy region of interest.

Substrate and thin layer materials have diverse band structure. Typically, electron confinement takes place close to band offsets (valence band top and conduction band bottom). The NG reduces DOS in both valence and conduction bands, and obtained results should be applied separately to each. For some material pairs, electron confinement can take place in more than two bands and also in subbands (case of large band offset). For each material pair, electron confinement energy regions should be determined carefully and $g(E,a)$ should be calculated separately (for all bands falling within the confinement regions [3]). It was mentioned in Sec. 1 that $g(E,a)$ is material (band structure) independent. Here, we clarify energy dependence of $g(E,a)$. NG strongly influences only eigenfunctions with low $k$ or large de Broglie wavelengths $2\pi/k$ that are comparable with NG dimensions ($a, w$). Eigenfunctions with higher $k$ values are only slightly influenced by the NG, and the changes in DOS for high $k$ and corresponding high $E$ values can be ignored. Consequently, $g(E,a)$ of a particular band depends on $E$ only for low $E$ values, corresponding (via $E(k)$ of a particular band) to first 5 (in our case) eigenvalues $k_n$ (n = 1, .., 5).

The values obtained for $\bar{g}$ suffice in the G-doping of solar-cell layers. As a previous investigation showed [3], values $\bar{g}$ =0.83-0.99 give the doping levels of $10^{18}$-$10^{19}$ cm$^{-3}$ required for multi-junction solar cells. The same or slightly higher G-doping levels suffice for ballistic transport devices [33]. However, such values of $\bar{g}_{min}$ are not sufficient to increase electron emission from the NG coatings of electrodes for room-temperature thermionic converters [5] and NG thermoelectric converters [4]; for these, $\bar{g}_{min}$ = of 0.1-0.2 is required. Further reduction in $\bar{g}_{min}$ is possible using double-sided NG layers with phase shifted gratings. This geometry has been used in an electromagnetic analogue [14] of our quantum system.

## 5. Conclusions

We numerically calculated the DOS in nanograting layers. The Method of Auxiliary Sources and MATLAB code were used to find the eigenvalues and eigenfunctions of the time-independent Schrödinger equation with Dirichlet boundary conditions. We found the first five eigenvalues and eigenfunctions for the NG layer and compared them with the corresponding eigenvalues and eigenfunctions of a planar layer. The eigenvalues increase, and for the higher $n$ values, they oscillate with increasing grating depth. The minima of the eigenfunctions appear along the entire area near the grating edge. We also calculated the correlation coefficients between the squares of the eigenfunctions (probabilities) and used them for probability normalization. Furthermore, the numerical data obtained were used to analyze the DOS. The NG was shown to reduce the probability of occupation of the quantum state. The integrated DOS reduces as much as 4.1 fold at $a$ =0.4. The minimum of the averaged probability mass function $\bar{g}_{min}$ =0.24 was found at $a$=0.4, which corresponds to a 4.1 -fold reduction in the DOS. Such a reduction leads to considerable changes in the electronic properties. In particular, such values for $\bar{g}_{min}$ are more than enough to obtain G-doping levels common for solar cells and ballistic transport devices.

**Acknowledgments**

The authors thank the Shota Rustaveli National Science Foundation (projects MTCU/84/3-250/13, DO/147/6-265/13 and FR/467/3-250/14) and the Science & Technology Center in the Ukraine (project STCU-5893) for providing funding. The work was also supported by the scholarship program of the German Academic Exchange Service (DAAD) and the Joint Research and Education Program, Forschungszentrum Jülich & Shota Rustaveli National Science Foundation (JS/6/6-265/13). An apparatus received from EU TEMPUS iCo-op project #530278- TEMPUS-1-2012-1-DE-TEMPUS-JPHES was used in the preparation of this work.